\documentstyle[prl,aps,multicol]{revtex}
\newenvironment{Fig}{
\begin{figure}[h]
\noindent\begin{minipage}[t]{3.4in}
\begin{center}
\leavevmode
\epsfxsize=3 true in
}
{
\end{center}
\end{minipage}
\end{figure}
}
\begin{document}
\bibliographystyle{prsty}
\input epsf
\title{1/f Noise in a Coulomb Glass}
\author{Clare C. Yu}
\address{
Department of Physics and Astronomy, University of California,
Irvine, Irvine, California 92697}
\date{\today}
\maketitle
\begin{abstract}
At low temperatures electron hopping in a three
dimensional Coulomb glass produces fluctuations in the single particle
density of states and hence in the resistivity. This results in a low 
frequency resisitivity noise spectrum which
goes as $f^{-\alpha}$ where $\alpha$ is very close to 1. 
This holds down to extremely low frequencies.
\end{abstract}


\begin{multicols}{2}
\narrowtext
Low frequency $1/f$ noise \cite{Dutta81,Weissman88,Kogan96}
is ubiquitous; it is found in a wide variety of conducting systems
such as metals, semiconductors, tunnel junctions \cite{Rogers84},
and even superconducting SQUIDs \cite{Koch83,Koelle99}.
Yet the microscopic mechanisms are still not well understood.
In some cases the electrons are in a 
Coulomb glass which is an insulator with randomly placed electrons
that have Coulomb interactions \cite{efrosbook}. Lightly doped semiconductors
and disordered metals are examples of such systems.
This paper focuses on $1/f$ noise in Coulomb glasses. 
Experimental studies on doped silicon inversion layers have shown that 
low frequency $1/f$ noise is produced by hopping conduction
\cite{Voss78}. More recent experiments have observed 1/f noise down
to 0.1 Hz in boron--doped silicon \cite{Massey97}.
Because the systems are glassy, electron hopping can occur on
very long time scales and this produces low frequency noise.
In this paper we show that the resulting noise spectrum goes
as $f^{-\alpha}$ where $f$ is frequency and the exponent 
$\alpha\approx 1$. 

Shklovski\u{i} developed the first theory of 1/f noise in Coulomb glasses.
He suggested that it is produced by fluctuations
in the number of electrons in an infinite percolating cluster
\cite{Shklovskii80}. These
fluctuations are caused by the slow exchange of electrons between
the infinite conducting cluster and small isolated donor clusters.
A more rigorous calculation combined with
numerical simulations \cite{Kogan81} of Shklovskii's model 
found a noise spectrum that went as $f^{-\alpha}$ where $\alpha$
was considerably lower than 1. Furthermore, below a minimum frequency
of order 1--100 Hz, the noise spectral density saturated and became a 
constant independent of frequency. A similar conclusion
holds for a model suggested by Kozub \cite{Kozub96} 
in which electron hops within finite clusters produce
fluctuations in the potential seen by hopping conduction electrons 
that contribute to the current. Kogan has argued that transitions between 
valleys in the energy landscape produces 1/f noise
because high barriers result in slow fluctuations in hopping conduction
\cite{Kogan98}.

In our approach electron hopping shifts the single particle energies
$\varepsilon$ because they depend on Coulomb interactions with other sites.
This leads to fluctuations in the single particle density of states 
$g(\varepsilon)$ which, in turn, produces
fluctuations in the conductivity. The conductivity depends on
the density of states $g(\varepsilon\approx \mu)$
in the vicinity of the Fermi energy $\mu$. Note that 
$g(\varepsilon\approx \mu)$ can be affected by hops between 
sites $i$ and $j$ even if
the energies on these sites are not near the Fermi
energy because an electron or hole on site $i$ or $j$ can
interact with other sites whose energy is (or was) near the Fermi energy.

We start with a model of the Coulomb glass that follows that 
of Baranovski\u{i}, Shklovski\u{i}, and \'{E}fros (BSE) \cite{Baranovskii80}.
In this model, the electrons occupy the sites of a periodic lattice,
and the number of electrons is half the number of sites.
Each site has a random onsite energy $\phi_{i}$ chosen from a uniform
distribution extending from $-A$ to $A$. Thus, $g_{o}$, the
density of states without interactions, is flat. A site can contain 0 or 1
electron. The Hamiltonian can be written as
\begin{equation}
H = \sum_{i}\phi_{i}n_{i} + \sum_{i>j}\frac{e^{2}}{\kappa r_{ij}}n_{i}n_{j}
\label{eq:hamiltonian}
\end{equation}
where the occupation number $n_{i}$ equals
$\frac{1}{2}$ if site $i$ is occupied
and $-\frac{1}{2}$ if site $i$ is unoccupied, $e$ is the electron charge,
$\kappa$ is the dielectric constant and $r_{ij}$ is the distance between
sites $i$ and $j$. The single site energy is $\varepsilon_{i}=\phi_{i}+\sum_{j}
\frac{e^{2}}{\kappa r_{ij}}n_{j}$.
At zero temperature Coulomb interactions between localized
electrons result in a so--called Coulomb gap in the single
particle density of states that is centered at the Fermi energy
\cite{efrosbook,Pollak70,Efros75}.

We will use Mott's argument for variable range
hopping \cite{efrosbook,Mott68,Ambegaokar71} to relate 
fluctuations in the density of states to fluctuations in the resistivity.
One can regard a Coulomb glass as a random resistor network \cite{Miller60}
with a transition between sites $i$ and $j$ associated with a resistance
$R_{ij}$ given by
\begin{equation}
R_{ij}=R_{ij}^{o}\exp(\xi_{ij})
\label{eq:resistanceij}
\end{equation}
where the prefactor $R_{ij}^{o}=kT/(e^{2}\gamma_{ij}^{o})$ with
$\gamma_{ij}^{o}$ given by \cite{efrosbook}
\begin{equation}
\gamma_{ij}^{o}=\frac{D^{2}|\Delta^{j}_{i}|}{\pi d s^{5}\hbar^{4}}
\left[\frac{2e^{2}}{3\kappa a}\right]^{2}\frac{r_{ij}^{2}}{a^2}
\left[1+\left(\frac{\Delta^{j}_{i}a}{2\hbar s}\right)^{2}\right]^{-4}
\label{eq:gamma0}
\end{equation}
where $D$ is the deformation potential, $s$ is the speed of sound,
$d$ is the mass density, and
$\Delta_{i}^{j} = \varepsilon_{j}-\varepsilon_{i}-e^{2}/\kappa r_{ij}$.
$\Delta_{i}^{j}$ is the
change in energy that results from hopping from $i$ to $j$.
$a=\kappa a_{B}$ is the effective Bohr radius of a
donor, and $a_{B}$ is
the usual Bohr radius ($a_{B}=\hbar^2/me^2$). We will set
the mass $m$ equal to the electron mass so that $a_{B}=0.529\AA$.
In eq. (\ref{eq:resistanceij}), the exponent is given by
\begin{equation}
\xi_{ij}=\frac{2r_{ij}}{a}+\frac{\varepsilon_{ij}}{kT}
\label{eq:xij}
\end{equation}
The exponent reflects the thermally activated hopping rate 
between $i$ and $j$ as well as the wavefunction overlap between the sites.
$\varepsilon_{ij}$ is given by \cite{efrosbook}:
\begin{equation}
\varepsilon_{ij}=\left\{ \begin{array}{ll}
|\varepsilon_j-\varepsilon_i|-\frac{e^2}{\kappa r_{ij}}, &
\mbox{$(\varepsilon_i-\mu)(\varepsilon_j-\mu)<0$}\\
{\rm max}\left[|\varepsilon_{i}-\mu|,|\varepsilon_{j}-\mu|\right], &
\mbox{$(\varepsilon_i-\mu)(\varepsilon_j-\mu)>0$}\\
\end{array}
\right.
\label{eq:Eij}
\end{equation}
At both high and low compensations, electron hopping usually occurs on
one side of the Fermi level $\mu$ and the lower expression applies. At 
intermediate compensations and in the regime of variable range hopping,
hopping electrons often cross the Fermi level and the upper expression
applies. 

In the regime of variable range hopping
Mott pointed out that hopping conduction at low temperatures comes from
states near the Fermi energy. Let $\tilde{\varepsilon}=\varepsilon-\mu$.
If we consider states within
$\varepsilon_{o}$ of the Fermi energy, then the concentration of states
in this band is 
$N(\varepsilon_{o})=\int^{\varepsilon_o}_{-\varepsilon_o}
g(\tilde{\varepsilon})d\tilde{\varepsilon}$ 
where $g(\tilde{\varepsilon})$ is the density of states measured from
the Fermi energy.
So the typical separation between sites is $R=[N(\varepsilon_{o})]^{-1/3}$.
To estimate the resistance corresponding to hopping 
between two typical states of the band, we replace $r_{ij}$ with $R$ and
$|\varepsilon_j-\varepsilon_i|$ with $\varepsilon_o$ in eqs. (\ref{eq:xij})
and (\ref{eq:Eij})
to obtain $\xi(\varepsilon_o)$. Minimizing $\xi(\varepsilon_o)$ yields
$\overline{\varepsilon}_o$. Plugging this into eqs. (\ref{eq:xij}) and
(\ref{eq:resistanceij}) yields the variable range hopping formula
for the resistivity 
$\overline{\rho}(T)=\rho_o(T)\exp(\xi(\overline{\varepsilon}_o))$. 

In our model the noise results from electron hopping 
which produces fluctuations in the density of states 
$g(\varepsilon)=\overline{g}(\varepsilon)+\delta g(\varepsilon)$, where
$\overline{g}(\varepsilon)$ is the average
density of states. This in turn creates fluctuations in $N(\varepsilon_o)$,
$\xi(\varepsilon_o)$, $\overline{\varepsilon}_o$, and $\rho(T)$. 
We can calculate these fluctuations by applying classical perturbation theory
\cite{Marion} to the derivation of the variable range formula. 
To first order, $\delta\xi(\varepsilon_o)=
\delta\rho(T)/\overline{\rho}(T)=-(2kTg(T,\overline{\varepsilon}_o))^{-1}
\int^{\overline{\varepsilon}_o}_{-\overline{\varepsilon}_o}
\delta g(T,\tilde{\varepsilon})d\tilde{\varepsilon}$. 
We have included
the temperature dependence of the density of states because 
at finite temperatures the Coulomb gap fills in and the density of
states no longer vanishes at the Fermi energy
\cite{Levin87,Mogilyanskii89,Grannan93,Vojta93,Li94,Sarvestani95}.
The autocorrelation function for the fluctuations in the resistivity is
\begin{eqnarray}
& &\frac{<\delta \rho(T,t_{2})\delta\rho(T,t_{1})>}{\overline{\rho}^{2}(T)}
=\frac{1}{4k^{2}T^{2}g^2(T,\overline{\varepsilon}_{o})} \nonumber\\
& &\int^{\overline{\varepsilon}_o}_{-\overline{\varepsilon}_o}
d\tilde{\varepsilon}
\int^{\overline{\varepsilon}_o}_{-\overline{\varepsilon}_o}
d\tilde{\varepsilon}^{\prime}
<\delta g(T,\tilde{\varepsilon},t_{2})\delta g(T,\tilde{\varepsilon}^{\prime},
t_{1})>
\label{eq:autocorr}
\end{eqnarray}
We assume that there is no correlation between the fluctuations
in the density of states at different energies, so 
\begin{eqnarray}
<\delta g(T,\tilde{\varepsilon},t_{2})\delta g(T,\tilde{\varepsilon}^{\prime},
t_{1})>&=&E<\delta g(T,\tilde{\varepsilon},t_{2})
\delta g(T,\tilde{\varepsilon},t_{1})>\nonumber\\
& & \delta(\tilde{\varepsilon}-\tilde{\varepsilon}^{\prime})
\label{eq:equalenergies}
\end{eqnarray}
where $E$ is an energy of order $2\overline{\varepsilon}_{o}$.
Furthermore we assume that the time and energy dependence of the density
of states autocorrelation function are separable, allowing us to write
\begin{equation}
\int^{\overline{\varepsilon}_o}_{-\overline{\varepsilon}_o}
d\tilde{\varepsilon}
<\delta g(T,\tilde{\varepsilon},t_{2})\delta g(T,\tilde{\varepsilon},t_{1})>
=C(\overline{\varepsilon}_o,T)f(T,t_2-t_1)
\label{eq:separable}
\end{equation}
where we are assuming translational invariance in time (stationary processes).
$C(\overline{\varepsilon}_o,T)$ is a function of 
$\overline{\varepsilon}_o$ and temperature. The function $f(T,t)$ characterizes
the time dependence of the return to equilibrium by the system
after it is perturbed by a fluctuation in the density of states. 
Inserting eqns.
(\ref{eq:equalenergies}) and (\ref{eq:separable}) in (\ref{eq:autocorr})
yields
\begin{equation}
\frac{<\delta \rho(T,t_{2})\delta\rho(T,t_{1})>}{\overline{\rho}^{2}(T)}
=\frac{E C(\overline{\varepsilon}_o,T)}
{4k^{2}T^{2}g^2(T,\overline{\varepsilon}_{o})} f(T,t_2-t_1)
\label{eq:rhorho}
\end{equation}
To relate this to the spectral density of the noise $S(\omega)$,
let $\psi_{\rho}(t_2-t_1)=<\delta \rho(T,t_2)\delta\rho(T,t_1)>$ and 
let $\psi_{\rho}(\omega)$ be the Fourier transform of $\psi_{\rho}(t_2-t_1)$.
According to the Wiener--Khintchine theorem \cite{Kogan96},
for a stationary process the spectral density of fluctuations
is given by \cite{comment:Eij}
\begin{eqnarray}
S_{\rho}(\omega)&=&2\psi_{\rho}(\omega) \nonumber\\
&=&\frac{E \overline{\rho}^{2}(T) C(\overline{\varepsilon}_o,T)}
{2k^{2}T^{2}g^2(T,\overline{\varepsilon}_{o})}f(T,\omega)
\label{eq:Sw}
\end{eqnarray}

We do not know the
temperature dependence of $f(T,t)$, so for the moment we will suppress this
and just refer to $f(t)$. Theoretical calculations find that after
large deviations from equilibrium, the density of states returns to equilibrium
with a time dependence given by $g(\mu,t)\sim -\ln t$ 
or $g(\mu,t)\sim t^{-\theta}$ where $\theta\ll 1$ \cite{Yu99}. This 
agrees with experiments done at low
temperatures \cite{Ovadyahu97,Vaknin98a,Martinez-Arizala98}. 
If we assume that these functional forms are also valid for $f(t)$
which applies to small perturbations, then we obtain $1/f$ noise.
We now describe the calculation leading to this conclusion 
\cite{Yu99}. One starts with the Hamiltonian (\ref{eq:hamiltonian})
but assumes that the Coulomb interactions are turned on at time $t=0$: 
\begin{equation}
H = \sum_{i}\phi_{i}n_{i} + \sum_{i>j}\frac{e^{2}}{\kappa r_{ij}}n_{i}n_{j}
\theta(t)
\label{eq:hamiltonianTime}
\end{equation}
where the step function $\theta(t)$ is 0 for $t<0$ and 1 for $t\geq 0$.
So for $t<0$ the noninteracting density of states is a constant $g_o$.
Once the interactions are turned on,
one follows the subsequent time development of the Coulomb gap.

The Coulomb gap arises because the
stability of the ground state with respect to single
electron hopping from an occupied site $i$ to an unoccupied
site $j$ requires \cite{efrosbook} $\Delta_{i}^{j}>0$.
So we need to subtract from the density of states
those states which violate this stability condition.
This leads to a self--consistent equation for the
density of states \cite{Baranovskii80,Yu99,Burin95}:
\begin{eqnarray}
g(\tilde{\varepsilon},& t)&= g_{o}  \prod_{j>i} \left( 1-a_{o}^{3}\int_{-A}^{A}
d\tilde{\varepsilon}^{\prime} g(\tilde{\varepsilon}^{\prime},t)
\theta(\frac{e^{2}}{\kappa r_{ij}}+\tilde{\varepsilon}-
\tilde{\varepsilon}^{\prime})\right.      \nonumber   \\
  &   & \left. \phantom{\int_{-A}^{A}}
  F(n_{i}^{\prime}=1,n_{j}^{\prime}=0)\theta (t-\tau_{ij}
  (\tilde{\varepsilon}^{\prime},\tilde{\varepsilon}, r_{ij}))\right)
\label{eq:dosprod}
\end{eqnarray}
where the single--site energy $\tilde{\varepsilon}_{i}=\tilde{\varepsilon}$, 
$\tilde{\varepsilon}_{j}=\tilde{\varepsilon}^{\prime}$, and
$a_{o}$ is the lattice constant. $n_{i}^{\prime}=n_{i}+1/2$; so
$n_{i}^{\prime}=1$ if site i is occupied and $0$ if site $i$ is unoccupied.
$F(n_{i}^{\prime},n_{j}^{\prime})$ 
is the probability
that donors $i$ and $j$ have occupation numbers $n_{i}^{\prime}$ and
$n_{j}^{\prime}$, respectively, while all other sites have their ground state
occupation numbers $\tilde{n}_{k}^{\prime}$. $\tau_{ij}^{-1}$ is the number
of electrons which jump from site $i$ to site $j$ per unit time.
$\theta(t-\tau_{ij})$ represents the fact that at time $t$, the
primary contributions to the change in the
density of states will be from those hops for which $\tau_{ij}<t$
\cite{comment:exp}. In writing eq. (\ref{eq:dosprod}), we
assume that these hops together with phonons
have equilibrated the system as much as is possible at time $t$.
The hopping rate $\tau_{ij}^{-1}$ is given by \cite{efrosbook}
\begin{equation}
\tau_{ij}^{-1}=\gamma_{ij}^{o}\exp(-\frac{2r_{ij}}{a})
[1+N(\Delta^{j}_{i})] F(n_{i}^{\prime}=1,n_{j}^{\prime}=0)
\label{eq:tau}
\end{equation}
where 
$N(\Delta^{j}_{i})$ is the phonon occupation factor and reflects
the contribution of phonon assisted hopping.  
We are also allowing for spontaneous
emission of phonons since we are considering a nonequilibrium situation
in which electrons hop in order to lower their energy. 
Following \cite{Baranovskii80,Yu99} 
we can rewrite the self--consistent equation $g(\tilde{\varepsilon},t)$: 
\begin{eqnarray}
& &g(\tilde{\varepsilon},t) =  g_{o}\exp \left\{-\frac{1}{2}\int_{-A}^{A}
d\tilde{\varepsilon}^{\prime}g(\tilde{\varepsilon}^{\prime},t) 
 \int_{a_{o}}^{\infty}dr 4\pi r^{2} \right. \nonumber       \\ 
& &\left. F(n(\tilde{\varepsilon})=1,n(\tilde{\varepsilon}^{\prime})=0)
\theta(\frac{e^{2}}{\kappa r}+\tilde{\varepsilon}-\tilde{\varepsilon}^{\prime})
\theta(t-\tau(\tilde{\varepsilon}^{\prime},\tilde{\varepsilon},r)) \right\} 
\label{eq:dos}
\end{eqnarray}
At low energies large distances play an important role and so we
have replaced the sum by an integral over $r$ in the exponent.
The origin is at site $i$.
$n(\tilde{\varepsilon})$ is the occupation probability of a site
with energy $\tilde{\varepsilon}$. 
$\tau(\tilde{\varepsilon}^{\prime},\tilde{\varepsilon},r)$
is given by (\ref{eq:tau}) with $r_{ij}$ replaced by $r$, 
$\tilde{\varepsilon}_{i}$ replaced by $\tilde{\varepsilon}$, 
and $\tilde{\varepsilon}_{j}$ replaced
by $\tilde{\varepsilon}^{\prime}$. 

Since it is not clear how the stability condition $\Delta_{i}^{j}>0$
can be applied to finite temperatures, we confine our calculations to
the case of $T=0$. In this case the phonon occupation factor 
$N(\Delta^{j}_{i})=0$ and the electron occupation factor $F(n_{i}=1,n_{j}=0)=1$ 
if $\tilde{\varepsilon}_i<0$ and $\tilde{\varepsilon}_j>0$. Otherwise 
$F(n_{i}=1,n_{j}=0)=0$. 
We can solve eq. (\ref{eq:dos}) iteratively on the computer. 
After a few iterations the typical difference between successive iterations
is typically less than 1 part in $10^{5}$. 
We find that the Coulomb gap develops slowly over many decades
in time \cite{Yu99}. 
After an infinite amount of time, the density of states at the
Fermi energy $\mu$ goes to zero and 
$g(\tilde{\varepsilon})\sim\tilde{\varepsilon}^2$.

The functional form of the time dependence of
$g(\tilde{\varepsilon},t)$ varies with the energy $\tilde{\varepsilon}$ and
with $g_o$. For conduction noise we are interested in the time dependence of
the density of states at the Fermi energy which is shown in Figure
\ref{fig:dosT0} for $10^{-8}$ s $< t < 10^{8}$ s.
For $g_{o}=2\times 10^{5}$ states/K--\AA$^{3}$, we can fit our results
to the form $g(\mu,t)=B_1\ln(t_o/t)$ 
where $t<t_o$, $t_o\cong 3\times 10^{43}$ sec, 
and for $g_{o}=6.25\times 10^{5}$ states/K--\AA$^{3}$,
$g(\mu,t)\cong B_2 t^{-\theta}$ where $\theta\cong 0.05$. 
The values of $B_1$, $B_2$, and the other parameters used to 
obtain these results are given in the caption of Figure \ref{fig:dosT0}. 
These fits change slightly for longer times. For example, for 
$10^{-8}$ s $< t < 10^{100}$ s and 
$g_{o}=2\times 10^{5}$ states/K--\AA$^{3}$, the fit to our results 
has the form $g(\mu,t)\sim t^{-\theta}$ where $\theta\cong 0.01$. Still 
we see that the density of states
at the Fermi energy approaches its equilibrium value
roughly logarithmically in
time. This is consistent with recent experiments on thin 
semiconducting \cite {Ovadyahu97,Vaknin98a} 
and metallic \cite{Martinez-Arizala98} films which have shown that the
system adjusted to changes in the Fermi energy approximately
logarithmically in time. These films were grown on insulating substrates
which separated them from a gate electrode that regulated the electron
density, and hence the chemical potential, of the film. 
The conductance was measured as a function
of the gate voltage. If the gate voltage was changed 
suddenly from, say, $V_{o}$ to $V_{1}$, the conductance had a very fast
initial rise, followed by a period of rapid relaxation, which in turn
was followed by a long period of very slow relaxation. 
The relaxation could be described by $\ln t$ or
$t^{-\theta}$ with $\theta$ being small and varying slowly with
time. This is consistent with our view
\cite{Yu99} that when the gate voltage is changed, the Fermi energy
changes, and time dependent relaxations arise because the system must dig
a new Coulomb gap in the density of states at the new Fermi energy.

So both theory and experiment indicate that
the nonequilibrium density of states approaches 
its equilibrium value roughly logarithmically in time. Returning
to the original model described by (\ref{eq:hamiltonian}), we assume that this
time dependence holds true in the linear response regime 
at low temperatures. If a 
fluctuation $\delta g(\mu,t=0)$ at $t=0$ pushes the density of states
away from its mean equilibrium value at the Fermi energy, then this
perturbation will decay according to $f(t)$
which enters into eqs. (\ref{eq:separable}) and (\ref{eq:rhorho}). 
Our nonequilibrium calculation indicates that $f(t)$ can have the form: 
\begin{equation}
f_1(t)=B_1\ln(\frac{t_o}{t})
\label{eq:ln}
\end{equation}
where $t<t_o$, and $t_o$ is on the order of the
age of the universe or longer, or
\begin{equation}
f_2(t)=B_2t^{-\theta}
\label{eq:power}
\end{equation}
where $\theta \ll 1$, and $B_1$ and $B_2$ are positive constants of order
$g_o$. In both cases $t$ is greater than some $t_{\rm min}$ of order
10$^{-8}$ s, say.
The time dependence is a function of the energy, so here we set
$\varepsilon=\mu$. Fourier transforming $f_1(t)$ and keeping the real part, 
we find that 
\begin{equation}
f_1(\omega)\approx\frac{\pi B_1}{2}\frac{1}{\omega}
\end{equation}
This implies that the noise spectral density $S(\omega)\sim 1/\omega$.
Fourier transforming $f_2(t)$ and keeping the real part yields
\begin{equation}
f_2(\omega)\approx \frac{\pi B_2\theta}{2}\frac{1}{\omega^{1-\theta}}
\end{equation}
for $\theta\ll 1$. This implies $S(\omega)\sim 1/\omega^{1-\theta}$.

To summarize, electron hopping leads to fluctuations in the density of states 
that relax back to equilibrium roughly logarithmically in time. This leads
to 1/f noise in the spectral density $S(\omega)$
of the noise in the resistivity. In particular we find that 
$S(\omega)\sim 1/\omega^{\alpha}$ where $\alpha=1$ 
if the relaxation is logarithmic in time, and $\alpha=1-\theta$
if the relaxation is a power law that goes as $t^{-\theta}$ 
where $\theta\ll 1$. In general 
$\alpha$ depends on temperature \cite{Massey97} and is weakly dependent on 
the noninteracting density of states $g_o$ and on the times scales.
As eq. (\ref{eq:Sw}) indicates, the noise amplitude also
depends on the temperature. Unfortunately we cannot ascertain these temperature
dependences because we do not know the temperature dependence of the
fluctuations $\delta g(T,\tilde{\varepsilon},t)$ in the density of states.
However we believe that our mechanism for 1/f noise should be valid
at low temperatures ($T\stackrel{<}{\sim} 20$ K)
where the logarithmic time dependence of the 
conductance is observed after the Coulomb glass has been pushed out
of equilibrium by the sudden application of a gate voltage 
\cite{Ovadyahu97,Vaknin98a,Martinez-Arizala98}.

I thank Herv\'{e} Carruzzo for helpful discussions.
This work was supported in part by ONR grant 
N00014-00-1-0005 and CULAR funds provided by the University of
California for the conduct of discretionary research by Los
Alamos National Laboratory.

\begin{Fig}
\epsfbox{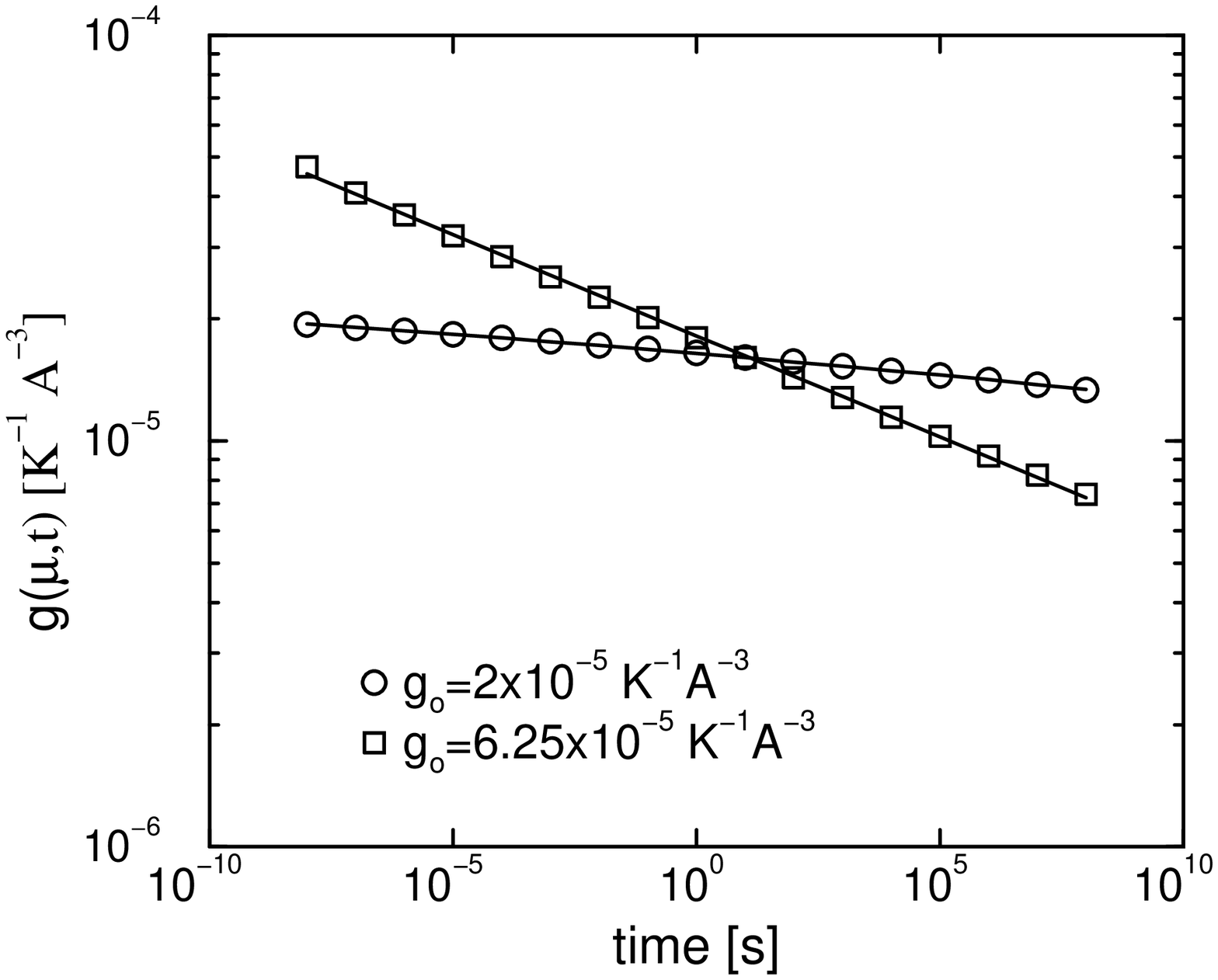}
\caption{Density of states $g(\varepsilon=\mu,t)$ at the Fermi energy
as a function of time for different values of $g_o$. Solid lines are fits
to the numerical integration of (\protect\ref{eq:dos}). The fit to
the $g_{o}=2\times 10^{5}$ states/K--\AA$^{3}$ data is given by
$B_1\ln(t_o/t)$ where $\ln t_o=100$ with time measured in seconds
and $B_1=1.64\times 10^{-7}$ states/K--\AA$^{3}$. The fit to the 
$g_{o}=6.25\times 10^{5}$ states/K--\AA$^{3}$ data is given by
$B_2t^{-\theta}$ where $\theta=0.050$ and 
$B_2=1.82\times 10^{-5}$ states/K--\AA$^{3}$.
Parameters used are $A=2 \times 10^4$ K, $\kappa=10$,
$d=7.18$ g/cm$^{3}$, $s=5.0\times 10^{5}$ cm/sec, $D=5 \times 10^{3}$ K,
and $a_{o}=4$ \AA. The density $d$ is chosen to be that of In$_{2}$O$_{3}$.
}
\label{fig:dosT0} 
\end{Fig}

\end{multicols}

\begin{thebibliography}{10}

\bibitem{Dutta81}
P. Dutta and P.~M. Horn, Rev. Mod. Phys. {\bf 53},  497  (1981).

\bibitem{Weissman88}
M.~B. Weissman, Rev. Mod. Phys. {\bf 60},  537  (1988).

\bibitem{Kogan96}
S. Kogan, {\em Electronic Noise and Fluctuations in Solids} (Cambridge
  University Press, Cambridge, 1996).

\bibitem{Rogers84}
C.~T. Rogers and R.~A. Buhrman, Phys. Rev. Lett. {\bf 53},  1272  (1984).

\bibitem{Koch83}
R.~H. Koch,  in {\em Noise in Physical Systems and 1/f Noise}, edited by M.
  Savelli, G. Lecoy, and J.-P. Nougier (Elsevier Science Pub., Amsterdam,
  1983), p.\ 377.

\bibitem{Koelle99}
D. Koelle {\it et~al.}, Rev. Mod. Phys. {\bf 71},  631  (1999).

\bibitem{efrosbook}
A. excellent review~is B.~I.~Shklovski\u{i} and A.~L. Efros, {\em Electronic
  Properties of Doped Semiconductors} (Spinger-Verlag, Berlin, 1984), and
  references therein.

\bibitem{Voss78}
R.~F. Voss, J. Phys. C {\bf 11},  L923  (1978).

\bibitem{Massey97}
J.~G. Massey and M. Lee, Phys. Rev. Lett. {\bf 79},  3986  (1997).

\bibitem{Shklovskii80}
B. Shklovski\u{i}, Sol. St. Comm. {\bf 33},  273  (1980).

\bibitem{Kogan81}
S.~M. Kogan and B.~I. Shklovksi\u{i}, Sov. Phys. Semicond. {\bf 15},  605
  (1981).

\bibitem{Kozub96}
V.~I. Kozub, Sol. St. Comm. {\bf 97},  843  (1996).

\bibitem{Kogan98}
S. Kogan, Phys. Rev. B {\bf 57},  9736  (1998).

\bibitem{Baranovskii80}
S.~D. Baranovski\u{i}, B.~I. Shklovski\u{i}, and A.~L. \'{E}fros, Sov. Phys.
  JETP {\bf 51},  199  (1980).

\bibitem{Pollak70}
M. Pollak, Disc. Faraday Soc. {\bf 50},  13  (1970).

\bibitem{Efros75}
A.~L. \'{E}fros and B.~I. Shklovski\u{i}, J. Phys. C {\bf 8},  L49  (1975).

\bibitem{Mott68}
N.~F. Mott, J. Non--Cryst. Solids {\bf 1},  1  (1968).

\bibitem{Ambegaokar71}
V. Ambegaokar, B.~I. Halperin, and J.~S. Langer, Phys. Rev. B {\bf 4},  2612
  (1971).

\bibitem{Miller60}
A. Miller and E. Abrahams, Phys. Rev. {\bf 120},  745  (1960).

\bibitem{Marion}
J.~B. Marion, {\em Classical Dynamics of Particles and Systems, 2nd ed.}
  (Academic Press, New York, 1970), p.\ 165.

\bibitem{Levin87}
E.~I. Levin, V.~L. Nguyen, B.~I. Shklovski\u{i}, and A.~L. \'{E}fros, Sov.
  Phys. JETP {\bf 65},  842  (1987).

\bibitem{Mogilyanskii89}
A.~A. Mogilyanski\u{i} and M.~E. Ra\u{i}kh, Sov. Phys. JETP {\bf 68},  1081
  (1989).

\bibitem{Grannan93}
E.~R. Grannan and C.~C. Yu, Phys. Rev. Lett. {\bf 71},  3335  (1993).

\bibitem{Vojta93}
T. Vojta, W. John, and M. Schreiber, J. Phys.: Condens. Matter {\bf 5},  4989
  (1993).

\bibitem{Li94}
Q. Li and P. Phillips, Phys. Rev. B {\bf 49},  10269  (1994).

\bibitem{Sarvestani95}
M. Sarvestani, M. Schreiber, and T. Vojta, Phys. Rev. B {\bf 52},  R3820
  (1995).

\bibitem{comment:Eij}
This equation for $S_{\rho}(\omega)$ is valid for both cases in eq.
  (\protect\ref{eq:Eij}).

\bibitem{Yu99}
C.~C. Yu, Phys. Rev. Lett. {\bf 82},  4074  (1999).

\bibitem{Ovadyahu97}
Z. Ovadyahu and M. Pollak, Phys. Rev. Lett. {\bf 79},  459  (1997).

\bibitem{Vaknin98a}
A. Vaknin, Z. Ovadyahu, and M. Pollak, Phys. Rev. Lett. {\bf 81},  669  (1998).

\bibitem{Martinez-Arizala98}
G. Martinez-Arizala {\it et~al.}, Phys. Rev. B {\bf 57},  R670  (1998).

\bibitem{Burin95}
A.~L. Burin, J. Low Temp. Phys. {\bf 100},  309  (1995).

\bibitem{comment:exp}
The exact form of the cutoff is not important; for example, replacing
  $\theta(t-\tau_{ij})$ with $[1-\exp(-t/\tau_{ij})]$ affects our results
  negligibly.

\end{thebibliography}
\end{document}